\documentclass[12pt]{article}
 \usepackage[nottoc]{tocbibind}
\usepackage[colorlinks=true,linkcolor=blue,urlcolor=magenta, citecolor=blue]{hyperref}
\usepackage{fullpage}
\usepackage{amssymb,graphicx}
\usepackage{amsmath}
\usepackage[margin=0.5cm]{caption}
\usepackage{hyperref} 
\usepackage{cite}
\usepackage{upgreek}

\def\d{\partial}
\def\l{\left(}
\def\r{\right)}

\newcommand{\be}{\begin{equation}}
\newcommand{\ee}{\end{equation}}
\newcommand{\bea}{\begin{eqnarray}}
\newcommand{\eea}{\end{eqnarray}}
\newcommand{\bg}{\begin{gather}}
\newcommand{\eg}{\end{gather}}
\newcommand{\bseq}{\begin{subequations}}
\newcommand{\eseq}{\end{subequations}}

\newcommand{\ket}[1]{| #1 \rangle}

\begin{document}
\baselineskip=15.5pt
\begin{titlepage}
\begin{center}
{\Large\bf  3D Yang--Mills Glueballs vs Closed Effective Strings }\\
\vspace{0.5cm}
{ \large
Sergei Dubovsky$^{a}$, Guzm\'an Hern\'andez-Chifflet$^{b}$, and Shahrzad Zare$^{a}$
}\\
\vspace{.45cm}
{\small  \textit{   $^a$Center for Cosmology and Particle Physics,\\ Department of Physics,
      New York University\\
      New York, NY, 10003, USA}}\\ 
      \vspace{.1cm}
      \vspace{.25cm}         
		{\small  \textit{   $^b$Instituto de F\'isica, Facultad de Ingenier\'ia,\\ Universidad de la Rep\'ublica,\\
				Montevideo, 11300, Uruguay}}\\ 
\end{center}
\begin{center}
\begin{abstract}


  Recent lattice results strongly support the Axionic String Ansatz (ASA) for quantum numbers of glueballs in 3D Yang--Mills theory. The ASA treats glueballs as closed bosonic strings. The corresponding worldsheet theory is a deformation of the minimal Nambu--Goto theory. In order to understand better the ASA strings and as a first step towards a perturbative calculation of the glueball mass splittings we compare the ASA spectrum to the  closed effective string theory. Namely, we model glueballs as excitations around the folded rotating rod solution with a large angular momentum $J$. The resulting spectrum agrees with the ASA in the regime of validity of the effective theory, {\it i.e.}, in the vicinity of the leading Regge trajectory. In particular, closed effective string theory correctly predicts that only glueballs of even spin $J$ show up at the leading Regge trajectory. Interestingly though, the closed effective string theory overestimates the number of glueball states far above the leading Regge trajectory. 

%
%
%
%
%
  
\end{abstract}
\end{center}
\end{titlepage}
\section{Introduction}
\label{sec:intro}
Gluodynamics in the 't Hooft large $N_c$ limit \cite{tHooft:1973alw} provides an example of a weakly coupled string theory, which is quite different from any well understood string model.
An Axionic String Ansatz (ASA) for the dynamics of long confining strings both at $D=3$ and $D=4$ spacetime dimensions was put forward in \cite{Dubovsky:2015zey}. This proposal led to a prediction (``the ASA spectrum") for the quantum numbers of short string states (glueballs) 
in $D=3$ Yang--Mills theory \cite{Dubovsky:2016cog}. At the time of the proposal the ASA spectrum was in agreement with the existing lattice data for 14 low lying glueball states corresponding to $N=0,1,2$ string levels \cite{Athenodorou:2016ebg}. Recent lattice results \cite{Conkey:2019blu} confirmed the ASA spectum for 25 heavier states at $N=3$ level. These results are also in a broad agreement with the ASA for 64 $N=4$ states, but more work needs to be done on the lattice side to test the ASA for these states.

We feel that these results provide a sound motivation to take the ASA proposal seriously and to understand it better. In particular, it is natural to ask whether it is possible to go beyond predicting the glueball quantum numbers, and to calculate also glueball mass splitings at each level. A natural setting to achieve this at least for some states is provided by the effective theory of rotating closed strings (see \cite{Hellerman:2013kba} for the treatment of the leading Regge trajectory in this framework and also \cite{Hellerman:2014cba,Sonnenschein:2020jbe,Sonnenschein:2018aqf,Sonnenschein:2015zaa} for other recent closely related work). Namely, one starts with a glueball of a large spin $J$ on the leading Regge trajectory ({\it i.e.}, a state with a  minimum energy $E$ in a sector with a fixed angular momentum $J$). At large $J$ this state may be described as a folded rotating string and other glueball states close to the leading Regge trajectory can be described by considering small perturbations around this background.

 In principle, one may then use the $1/J$ expansion to calculate perturbatively  the masses of glueball states in the vicinity of the leading Regge trajectory. There is a technical difficulty though in implementing this program due to the presence of a fold singularity for the classical rotating rod solution at $D=3$\footnote{The same difficulty arises also at $D=4$. At $D>4$ the singularity can be avoided by considering a string rotating in two orthogonal planes \cite{Hellerman:2013kba}.}. In the current paper we will not attempt to address this difficulty and will pursue a more modest goal. Namely, we will study glueball quantum numbers coming out from the effective string theory and compare them to the ASA spectrum.
 
 We start in section~\ref{sec:ASA} with a brief review of the ASA spectrum. To make the paper self-contained we present there a derivation of the ASA spectrum. This derivation is slightly different from the one presented in  \cite{Dubovsky:2016cog}, and it is instructive to compare it with the later 
closed  effective string theory calculation. The latter is presented in section~\ref{sec:effective}. In section~\ref{sec:comparison} we compare the two spectra. We present our conclusions and discuss future directions in section~\ref{sec:concl}.

\section{The ASA recap}
\label{sec:ASA}
There are two major ingredients entering into the derivation of the  ASA glueball spectrum. First, it is assumed that the glueball Hilbert space ${\cal H}_{gl}$ can be decomposed into a sum over string levels labeled by $N$,
\be
\label{levelsum}
{\cal H}_{gl}=\sum_{N=0}^\infty{\cal H}_L\otimes {\cal H_R}\;.
\ee
Here ${\cal H}_L$ and ${\cal H}_R$ are Hilbert spaces of left- and right-moving excitations on the string worldsheet. For confining strings one expects to find
\[
{\cal H}_L={\cal H}_R
\] 
as a consequence of the charge conjugation $C$, which acts by exchanging left- and right-movers
\[
C(\psi\otimes\chi)=\chi\otimes\psi\;.
\]

The decomposition (\ref{levelsum}) may be thought of as a part of a definition of what it means for glueballs to be closed string states without any additional massive degrees of freedom on the worldsheet. Note that if the latter were present, the decomposition  (\ref{levelsum}) would not hold due to a possibility to add the corresponding massive excitations at rest.
 
 Of course, for the decomposition (\ref{levelsum}) to be useful in practice, one needs to make some assumptions about the masses of different states. For critical strings all states at the same level are completely degenerate.
 This degeneracy can be traced to the integrability of the worldsheet  theory in the critical case.  
 The worldsheet theory of confining strings is non-integrable. On the other hand, one of the major motivations for the ASA is the idea of approximate integrability on the worldsheet of confining strings. 
Empirically, the approximate integrability is suggested by a certain intriguing numerological coincidence in the coupling of the worldsheet axion as extracted from the lattice data \cite{Dubovsky:2013gi,Dubovsky:2015zey}. A closely related coincidence was revealed also by applying the $S$-matrix bootstrap  to the fluxtube dynamics \cite{EliasMiro:2019kyf}.
 
 Coming from a theory side,  low energy  integrability arises as a consequence of the tree level integrability of the Nambu--Goto theory \cite{Dubovsky:2012wk}. At high energies integrability can be understood as a byproduct of asymptotic freedom and confinement \cite{Dubovsky:2018vde}.
 Namely, it is natural to identify high energy degrees of freedom on the worldsheet with partons. Asymptotic freedom implies the absence of (hard) particle production in high energy parton scattering. The presence of a confining string implies that there is a non-trivial phase shift in the worldsheet scattering. Hence, the worldsheet theory is forced to turn integrable at high energies rather than to reduce to just a free one\footnote{Note that one should be careful not to take these arguments too literally. Asymptotic freedom does not prevent soft particle production. Hence integrability is expected to be found only for a certain subset of suitably defined inclusive hard observables.}.
 
 Motivated by these considerations the ASA assumes that glueball states corresponding to the same string level are approximately degenerate. This is a crucial assumption which makes the decomposition (\ref{levelsum}) practically useful.
 
 The second major step in the derivation of the ASA spectrum relies on the assumption that left- and right-mover's Hilbert spaces ${\cal H}_{L,R}$ are the same as the Hilbert space of open strings ${\cal H}_{open}$,
 \be
 \label{LRopen}
 {\cal H}_{L}={\cal H}_{R}={\cal H}_{open}\;.
 \ee
 Unlike the tensor product decomposition (\ref{levelsum}) we are not aware how to justify (\ref{LRopen})
on general grounds. The main motivation for (\ref{LRopen}) is that it holds for critical strings. 

Once this assumption is accepted, one may explore ${\cal H}_{L,R}$ perturbatively.  Namely, glueball states at the leading Regge trajectory ({\it i.e.}, minima of energy $E$ at fixed angular momentum $J$) at large $J$ can be described by 
the classical  rotating rod solution. Low lying excitations above the  leading Regge trajectory can be described then by quantizing small perturbations around the rotating rod.

Let us see how this works in practice. Our analysis here is similar to the one presented in  \cite{Dubovsky:2016cog}. A minor technical difference is that we are not restricting to the subspace of fixed $J$ as was done in \cite{Dubovsky:2016cog}, but consider all possible perturbations. 
  
  Dynamics of a long smooth string is described by the Nambu--Goto action,
\be
\label{NG}
S_{NG}=-\ell_s^{-2}\int d\tau d\sigma\sqrt{-\det \d_\alpha X^\mu\d_\beta X_\mu}+\dots\;,
\ee 
where dots stand for higher derivative terms, which will be ignored in our analysis. In principle, one may work directly with this action. However, as is often the case, it is more convenient to work with the equivalent Polyakov action
\be
\label{Pol}
S_{Pol}=-\ell_s^{-2}\int d\tau d\sigma \sqrt{-h}{1\over 2}h^{\alpha\beta}\d_\alpha X^\mu\d_\beta X_\mu+S_{PS}+\dots\;.
\ee
Here $S_{PS}$ is the Polchinski-Strominger (PS) term \cite{Polchinski:1991ax} which has to be introduced given that we work in a non-critical space-time dimension (see \cite{Hellerman:2014cba} for a nice modern introduction into the PS formalism).  This term is important for calculating next-to-leading (one loop) effects in the $1/J$ expansion. For the purpose of the leading order (tree level) analysis presented here it can be ignored. 

For open strings we choose the following range of the $\sigma$ coordinate 
\[
\sigma\in [0,\pi]\;.
\]
 At the
string  end-points the embedding coordinates satisfy  the Neumann boundary conditions,
\be
\label{Neum}
\sqrt{-h}h^{\sigma\alpha}\d_\alpha X^\mu(0)=\sqrt{-h}h^{\sigma\alpha}\d_\alpha X^\mu(\pi)=0\;.
\ee 
In general, these get corrected due to higher order dotted terms in (\ref{NG}), (\ref{Pol}), which include also boundary localized contributions such as the end-point mass. Again, these effects can be ignored at the leading order in the $1/J$ expansion.

\begin{figure}
	\centering
	\includegraphics[width=0.5\linewidth]{"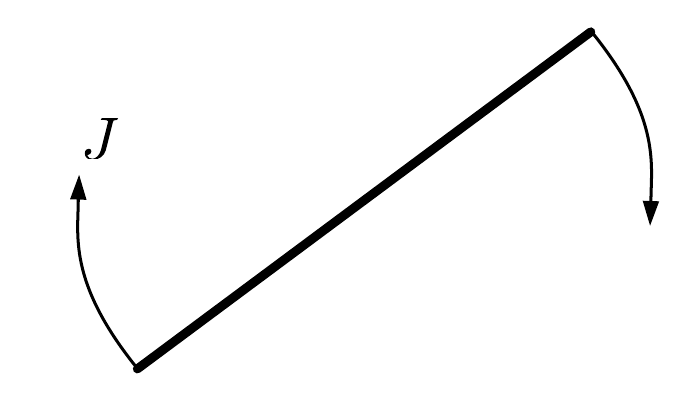"}
	\caption{Open string rotating with angular momentum J.}
	\label{fig:rod}
\end{figure}

The classical rotating rod solution of an energy $E_r$ and an angular momentum $J_r$ takes then the following form (see Fig.~\ref{fig:rod})
\begin{gather}
\label{X0}
X_{r}^0=\sqrt{2J_r\over \pi}{\ell_s}\tau\;\\
\label{X}
X_{r}\equiv X_{r}^1+iX_{r}^2=\sqrt{2J_r\over \pi} {\ell_s} e^{i\tau}\cos\sigma\;.
\end{gather}
The corresponding Polyakov metric is flat 
\be
\label{gauge}
h_{\alpha\beta}=\eta_{\alpha\beta}\;.
\ee
The energy and the angular momentum of this solution are related by the classical Regge formula,
\be
\label{Regge}
E_r^2={2\pi J_r\over \ell_s^2}\;.
\ee

To quantize small perturbations around this solution, we will make use of  the first order formalism (which actually precedes the Polyakov 
description and goes back to \cite{Goddard:1973qh}; its concise pedagogical exposition can be found, e.g., in  \cite{Brink:1988nh,Gleb}). In this approach one replaces the leading Polyakov action with the equivalent first order action
\be
\label{fo}
S_{first\;order}=\int d\tau d\sigma \l \Pi_\mu\d_\tau X^\mu+ {1\over 2\sqrt{-h}h^{\tau\tau}}\l \ell_s^2\Pi^\mu \Pi_\mu+\ell_s^{-2}\d_\sigma X^\mu
\d_\sigma X_\mu\r+{h^{\tau\sigma}\over h^{\tau\tau}}\Pi_\mu\d_\sigma X^\mu\r\;.
\ee
This rewriting makes it clear that the classical Nambu--Goto dynamics is governed by a trivial Hamiltonian and two Virasoro constraints.
The latter are first class constraints, {\it i.e} these are gauge symmetry generators.
 The advantage of the first order formalism is that it makes this canonical structure manifest.

In this formalism, it is common to use the light cone gauge for the analysis of the string spectrum. However, in the context of $1/J$ expansion, the so-called static gauge is more convenient.  Details of the gauge fixing procedure can be found in Appendix~A, here we present the results only.
Analogously to the light cone gauge, as a first gauge condition
we fix $X^0$ in the form (\ref{X0}).
As a second gauge fixing condition we impose
\be
\label{P0}
\Pi_0=-{E(\tau)\over \pi}\;.
\ee
Here $E(\tau)$ is the target space energy of the string, which stays constant on-shell. 

Then, by varying the action (\ref{fo}) w.r.t. $\Pi_0$ one arrives at
\be
\label{h00}
h^{\tau\tau}\sqrt{-h}=-{E\over E_r}\;.
\ee
On the other hand, variation of (\ref{fo}) w.r.t. $X^0$ gives
\be
\label{h01}
\d_\sigma\l{h^{\tau\sigma}\over h^{\sigma\sigma}}\r=-{\d_\tau E\over E}\;.
\ee
By making use of the Weyl symmetry we may set
\[
\sqrt{-h}=1\;.
\]
Then (\ref{h00}) and (\ref{h01}) imply that all components of $h^{\alpha\beta}$ are independent of $\sigma$ and the Neumann boundary condition (\ref{Neum}) enforces
\[
h^{\tau\sigma}=0\;.
\]
Thus, similarly to the treatment of the light cone gauge presented in \cite{Gleb}, in the first order formalism the condition analogous to
(\ref{gauge}) comes about as a result of solving for the non-dynamical Lagrange multipliers, rather than
from direct  conformal gauge fixing as in the conventional Polyakov treatment.  

A general perturbation of the rotating rod solution can be parametrized in the following way in this gauge
\be
\label{genX}
X=e^{i\delta(\tau)}X_r+ e^{i(\tau+\delta(\tau))}\l x(\tau) +  \sqrt{2\over\pi}\sum_{n=1}^\infty \chi_n(\tau)\cos n\sigma \r \;.
\ee
This parametrization is one-to-one provided $\delta$ and $\chi_1$ are real, $x$ and $\chi_n$'s with $n>1$ are complex and
\be
\label{poscond}
\sqrt{2J_r\over \pi}{\ell_s}+\chi_1>0\;.
\ee
In this parametrization, we separated the phase  $\delta$, which may be thought of as a  Goldstone mode arising as a result of
the spontaneous breaking of time translations and spatial rotations down to the diagonal subgroup in the presence of the stationary classical background (\ref{X0}), (\ref{X}). The unbroken diagonal combination ensures
that the action for perturbations enjoys time translation invariance $\tau\to \tau+const$. 
The effective theory around the leading Regge trajectory treats all excitations perturbatively. Only the Goldstone phase $\delta$ and the average string position $x(\tau)$ are not assumed to be small.

In the first order formalism, the momentum 
\[
\Pi=\Pi_1+i \Pi_2
\]
enters as an independent field.
Its classical background is given by
\[
\Pi_{r}=\ell_s^{-2}\d_\tau X_r=i\ell_s^{-2}X_{r}\;.
\]
The momentum field of a general perturbed rotating string can be parametrized as
\be
\label{genP}
\Pi=e^{i\delta(\tau)}\Pi_r+ e^{i(\tau+\delta(\tau))} \l \frac{p(\tau)}{\pi}+ \sqrt{\frac{2}{\pi}}\sum_{n=1}^\infty \pi_n(\tau)\cos n\sigma \r,
\ee
where $p$
and $\pi_n$'s are all complex.

Note that in this parametrization the total physical momentum of the string is given by
\be
p_{ph}(\tau)=e^{i(\tau+\delta(\tau))}p(\tau)
\ee
rather than by $p(\tau)$. Also, as a consequence of Lorentz symmetry, it should be possible to run the effective field theory analysis without assuming that $p_{ph}$ is small. However, for simplicity, we 
will not attempt to do it here.

In this parametrization, the canonical part of the action (\ref{fo}) ({\it i.e.}, the part not including the Virasoro constraints)
takes the form
\be
\label{S0}
S_0=
\int d\tau\l\Omega-\ell_s\sqrt{2J_r\over \pi}E+J\r\;,
\ee
where 
\be
\label{canonical}
\Omega=\mbox{\rm Re}\l p^*\d_\tau x+\sum_{n=2}^\infty \pi_n^*\d_\tau\chi_n\r+\d_\tau \chi_1\mbox{\rm Re}\,\pi_1+J\d_\tau\delta 
\ee
is the canonical one-form, and
\be
\label{J}
J= \mbox{\rm Re} \l ip^*x + J_r+\sqrt{J_r\over 2} \l\ell_s^{-1}\chi_1+i\ell_s\pi_1^*\r +i\sum_{n=1}^\infty \pi_n^*\chi_n\r
\ee
is the angular momentum of the string.

The remaining step is to impose the Virasoro constraints. As a result of the gauge fixing, these are second class constraints now, and can be imposed simply by eliminating some of the variables. Before implementing this step in practice, notice that both Virasoro constraints are independent of the phase variable $\delta$. Hence, from the form of the action (\ref{S0}) we can conclude right away that the phase $\delta$ is canonically conjugate to the angular momentum,
\be
\label{pid}
\pi_{\delta}=J\;,
\ee
so that $J$ generates shift symmetry of $\delta$. Given that $\delta$ is a $2\pi$-periodic variable, $\delta\sim \delta+2\pi$, this implies that upon quantization $J$ may take integer values only,
\be
\label{Jquant}
J\in \mathbf{Z}\;.
\ee
Note that up to now our discussion was not relying on any perturbative expansion. Hence, the quantization condition (\ref{Jquant}) holds unchanged in the full quantum theory. Of course, this is expected on general grounds for a bosonic Poincar\'e invariant theory.

On the other hand, the light cone gauge quantization of $D=3$ bosonic strings leads to irrational anyonic states in the spectrum
\cite{Mezincescu:2010yp} (the same applies also to $D=3$ superstrings \cite{Mezincescu:2011nh}). We conclude that the light cone gauge quantization is anomalous for $D=3$ (super)strings. Unlike for other non-critical dimensions, at $D=3$ this is a global anomaly, which does not show up in the local Poincar\'e algebra. Most likely it is related to the use of the singular string configurations in the light cone gauge.
Adding these configurations distort the proper geometrical structure of the physical phase space responsible for the quantization condition (\ref{Jquant}).

Let us solve the Virasoro constraints now. This looks hard to do exactly in the gauge we are using, so at this stage we will resort to the perturbative $1/J_r$ expansion. Details of solving the linearized Virasoro constraints are provided in the Appendix B. After the dust settles, the result is that the Hilbert space of open string perturbations around the classical background is generated by a single tower of oscillator operators $a^\dagger_n $ with $n$ going from $1$ to $\infty$, acting on states of fixed internal angular momentum $\ket{ I}$. Here the internal angular momentum $I$ is  related to the total angular momentum $J$ by (\ref{internalI}). Both $I$ and $J$ are conserved, and take integer values.
All creation  operators $a_n^\dagger$ commute with both $I$ and $J$,  so their values are unchanged under the action of  $a_n^\dagger$.

At the leading order in the $1/J$ expansion, the energy is found to be
\begin{equation}
E^2 = |p|^2 + \frac{2\pi}{\ell_s^2}N
\label{energy_vs_N_open}
\end{equation}
where we introduced the level operator $N$ given by
\begin{equation}
\label{level}
N = I + \sum_{n=1}^{\infty}(n+1)a_n^\dagger a_n\;.
\end{equation}
These results agree with those obtained in \cite{Dubovsky:2016cog},  where the quantization was performed in the string rest frame and in the sector with fixed value of $I$. The canonical treatment of the theory 
emphasized in the approach presented here, allows us also to derive the quantization condition for $I$ (and $J$).

The perturbative spectrum (\ref{energy_vs_N_open}) is accurate for sufficiently low lying excitations above the leading Regge trajectory with $J_r\gg1$ and $I$ close to $J_r$. In particular,  (\ref{energy_vs_N_open}) provides an exact description of the fixed $I$  sector in the limit $I\to \infty$. Of course, the same analysis can be applied starting with a rotating rod solution with a negative angular momentum which would lead us to the same spectrum (\ref{energy_vs_N_open}) at large negative angular momentum $I$, with $I$ replaced by $|I|$ in (\ref{level}).

Clearly, the $1/J$ expansion breaks down at $J\sim 1$. Perhaps the most controversial step in the ASA is an assumption that for state counting one may use  (\ref{energy_vs_N_open}) for small values of $I$, including even non-rotating $I=0$ string states.
  The ASA closed string Hilbert space is constructed by plugging  $\mathcal{H}_{open}$ obtained in this way into the tensor square decomposition  \eqref{levelsum}. Remarkably, as we review in section~\ref{sec:comparison}, the resulting closed string spectrum 
agrees with the glueball spectrum observed in lattice simulations.  
  
%
%
%
%
%

\section{The spectrum of   closed effective  strings}
\label{sec:effective}
As reviewed in the previous section, the ASA spectrum of closed strings is obtained by perturbative quantization of  \emph{open strings} around a rotating rod background and then using the resulting Hilbert space to build the closed string Hilbert space. However, it is also possible to perform this perturbative quantizaton procedure directly for closed strings. As we will see, the resulting Hilbert space and spectra  differ from  the ASA results, although they coincide close to the leading Regge trajectory.  In this section we work out the perturbative quantization procedure for closed strings.

\begin{figure}
	\centering
	\includegraphics[width=0.5\linewidth]{"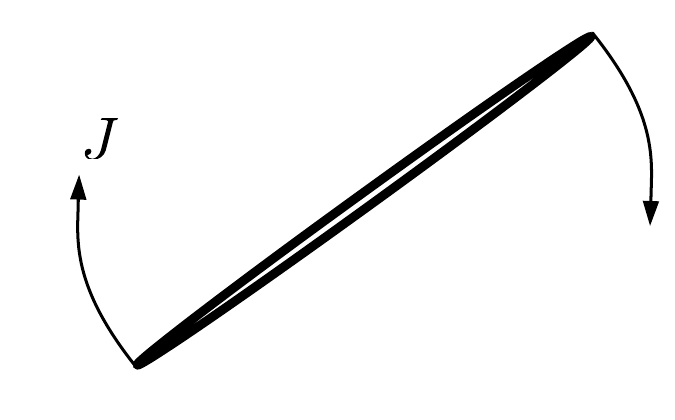"}
	\caption{Folded string rotating with angular momentum $J$.}
	\label{fig:closed-string}
\end{figure}

 Quantization of small perturbations around the classical rotating rod background for closed strings proceeds much in the same way as for open strings. However,  as will be discussed shortly, one needs to account for some important subtleties with the gauge fixing.
 The classical closed string rotating rod solution of energy $E_r$ and angular momentum $J_r$ can be written in the form (see Fig.~\ref{fig:closed-string})
\begin{gather}
\label{X0Closed}
X_{r}^0=\sqrt{J_r\over \pi}{\ell_s}\tau\;\\
\label{XClosed}
X_{r}\equiv X_{r}^1+iX_{r}^2=\sqrt{J_r\over \pi} {\ell_s} e^{i\tau}\cos\sigma\;.
\end{gather}
where the range of the $\sigma$ coordinate is now given by
\[
\sigma\in [0,2\pi]\;.
\]
The corresponding classical Regge formula relating  $J_r$ and $E_r$ takes form
\be
\label{ReggeClosed}
E_r^2={4\pi J_r\over \ell_s^2}\;.
\ee
As for open strings, in the static gauge $X^0$ is fixed in the form (\ref{X0Closed}) and its conjugate momentum $\Pi_0$ is fixed to be
\be
\label{P0Closed}
\Pi_0=-{E(\tau)\over 2\pi}\;.
\ee
which differs from its open string counterpart \eqref{P0} only by a multiplicative constant prefactor.

 Importantly, unlike  for open strings, this prescription does not fully fix  all the available gauge freedom. Indeed, in the notations of Appendix A, gauge transformations with 
 \[
 \xi^{\tau}=0
 \] 
 and $\xi^{\sigma}$ being any $\sigma$-independent function of $\tau$ are compatible with static  gauge conditions. These gauge transformations are absent for open strings. Indeed, $\xi^{\sigma}$ must vanish at the string endpoints for the transformed $X^\mu$ to satisfy Neumann boundary conditions. 
However, these transformations are allowed for closed strings. 
This can also be understood by noting 
 that on-shell these gauge transformations correspond to constant time-dependent shifts of $\sigma$ which are allowed transformations for closed strings.

 To take care of this residual gauge freedom, we parametrize a general perturbation of our classical background as
\be
\label{genXClosed}
X=e^{i\delta(\tau)}X_r+ e^{i(\tau+\delta(\tau))} \l x(\tau) +\sqrt{1\over 2\pi}\sum_{n\neq -1,0} \chi_n(\tau)e^{in\sigma} \r +\sqrt{1\over 2\pi} e^{i(\tau+\delta'(\tau))}\chi_{-1}(\tau)e^{-i\sigma}\;.
\ee
For the phases $\delta$ and $\delta'$ to be well defined we require $\chi_{\pm1}$  to be real as well as to satisfy the positivity conditions
\begin{gather}
\sqrt{J_r\over 2}{\ell_s}+\chi_{\pm1}>0\label{poscond1}\end{gather}
 We may now fix the residual gauge freedom by imposing the condition
\begin{equation}
\delta = \delta'.
\end{equation}
Note however that this condition still leaves a  discrete residual gauge freedom to shift  $\sigma$ by $\pi$. At this stage we will keep this remaining $Z_2$ subgroup  unfixed.
It plays an important role later.

 The canonical momentum field of the rotating closed string
\[
\Pi=\Pi_1+i \Pi_2
\]
developes a classical background  given by
\[
\Pi_{r}=\ell_s^{-2}\d_\tau X_r=i\ell_s^{-2}X_{r}\;.
\]
The momentum field of a general closed string configuration is parametrized much in the same way as for open strings
\be
\label{genPclosed}
\Pi=e^{i\delta(\tau)}\Pi_r + e^{i(\tau+\delta(\tau))} \l \frac{p(\tau)}{2\pi}+ \sqrt{1\over 2\pi}\sum_{n\neq 0} \pi_n(\tau)e^{in\sigma}\r \,,
\ee
where $p$ and  $\pi_n$'s are all complex. The canonical part of the action is  the same as \eqref{S0}, where now the symplectic one-form is
\be
\label{canonicalClosed}
\Omega=\mbox{\rm Re}\l p^*\d_\tau x+\sum_{\substack{n\neq 0, \pm1}} \pi_n^*\d_\tau\chi_n\r +\d_\tau \chi_1\mbox{\rm Re}\,\pi_1+\d_\tau \chi_{-1}\mbox{\rm Re}\,\pi_{-1}+J\d_\tau\delta 
\ee
and the total angular momentum is given by
\be
\label{JClosed}
J=
\mbox{\rm Re} \l ip^*x + J_r+\sqrt{J_r\over 2} \l\ell_s^{-1}(\chi_1 +\chi_{-1})+i\ell_s(\pi_1^* +\pi_{-1}^*)\r +i\sum_{\substack{n\neq 0}} \pi_n^*\chi_n\r.
\ee

In this setup, we now perform a procedure completely analogous to the one followed for open strings and solve the Virasoro constraints perturbatively in $1/J_r$. The details are worked out in Appendix C. The end result is that we arrive at  the ``almost physical" perturbative Hilbert space constructed out of two towers of oscillators $a_n$ and $b_n$. As before these commute with the total and internal angular momentum operators $J$ and $I$. Thus the ``almost physical" perturbative Hilbert space  is generated by acting with the creation operators $a_n^{\dagger}$ and $b_n^{\dagger}$ on the internal momentum eigenstates $\ket{I}$.

To the leading order in the $1/J$ expansion the energies of the closed string states are given by
\begin{equation}
E^2 = |p|^2 + \frac{4\pi}{\ell_s^2}N
\label{energy_vs_N_closed}
\end{equation} 
where the number operator $N$ is 
\begin{equation}
\label{Nab}
N = I + \sum_{n=1}^{\infty}\left((n+1)a_n^\dagger a_n+ (n+1)b_n^\dagger b_n \right).
\end{equation}
As the name implies,  the ``almost physical" Hilbert space is not yet the physical Hilbert space of effective strings. The reason is the presence of the residual $Z_2$ gauge symmetry, 
\be
\label{Z2}
\sigma\to\sigma+\pi\;.
\ee
This implies that to construct the physical Hilbert space one needs to project out  states which are not invariant under (\ref{Z2}).
By inspecting the $n=\pm1$ sector, one concludes that this $Z_2$ transformation corresponds to the shift of the phase $\delta$
\begin{equation}
\delta \rightarrow \delta + \pi\;.
\end{equation}
Then one finds that the $a_n$ and $b_n$ oscillators transform as
%
%
%
\begin{align}
a_m \rightarrow a_m e^{i(m+1)\pi}\\
b_m \rightarrow b_m e^{i(m+1)\pi}.
\end{align}
On the other hand, the trasformation rule for $\delta$ implies that eigenstates of $I$ transform as\footnote{Notice that in the parametrization given by \eqref{genXClosed} and \eqref{genPclosed} both $x$ and $p$ get a phase under the $Z_2$ transformation and $J$ is the canonically conjugate variable to $\delta_c$, as defined in Appendix B, which transforms under $Z_2$ in the same way as $\delta$. However, upon implementing transformations \eqref{transfphysx} and \eqref{transfphysp}, $x$ and $p$ no longer transform under $Z_2$. In these variables $I$ is canonically conjugate to $\delta_c$ and equation \eqref{phaseI} follows.}
\begin{equation}
\ket{I}\rightarrow e^{i{I}\pi}\ket{I}.
\label{phaseI}
\end{equation} 
Therefore, if we take a state $\ket{\Psi}$ in our ``almost physical" Hilbert space
\begin{equation}
\ket{\Psi} = \prod_{k\geq 1}(a_k^\dagger)^{n_k}\prod_{j\geq 1}(b_j^\dagger)^{m_j}\ket{I},
\end{equation}
it transforms under $Z_2$ as
\begin{equation}
\ket{\Psi} \rightarrow e^{iN\pi}\ket{\Psi}.
\end{equation}
Hence the physical Hilbert space is built up of even $N$ states only. In particular, this implies that all physical states on the leading Regge trajectory carry even spin $I$ in agreement with the lattice data. It is interesting that this result comes out in the ASA spectrum in a very different way---it arises there as an immediate consequence of the tensor square structure.
 Let us now perform a detailed comparison of the ASA spectrum, effective closed strings and the lattice glueball data.

\section{Comparison between the ASA and effective string theory spectra}
\label{sec:comparison}
It is often convenient to package numbers of string states $P_{N,J}$ with given level $N$ and spin $J$\footnote{Here we work in the rest frame, $p=0$, so that there is no distinction between $J$ and $I$.} into a generating function defined as
\be
\chi(x,\theta)=\sum_{N,J}x^Ne^{iJ\theta}P_{N,J}\;.
\ee
For instance, the generating function for the ASA open string spectrum derived in section ~\ref{sec:ASA} takes the following form \cite{Dubovsky:2016cog}
\be
\label{chiopen}
\chi_{open}(x,\theta)=
\sum_{N=0}^\infty x^N\chi_N(\theta)
={(1-x)(1-x^2)P(x)\over 1+x^2-2x\cos\theta}\;,
\ee
where $P(x)$ is the Euler partition function
\begin{equation}
P(x) = \prod_{m=1}^{\infty}(1-x^m)^{-1} = \sum_{m=0}^{\infty}x^mP_m\;.
\end{equation}
Then the generating function for the closed string ASA spectrum is given by
\be
\label{chiASA}
\chi_{cl}(x,\theta)=\sum_{N_c=0}^\infty x^{2N_c}\chi_{N_c}(\theta)^2\;.
\ee
On the other hand, it is straightforward to see that the generating functional for the closed effective string spectrum takes the following form before the $Z_2$ projection
\be
\label{chieff}
\chi_{eff}(x,\theta)={(1-x)^2(1-x^2)P(x)^2\over 1+x^2-2x\cos\theta}\;.
\ee
As discussed in section~\ref{sec:effective}, the physical closed effective string spectrum is obtained as a result of the $Z_2$ projection, which amounts to keeping only even powers of $x$ in (\ref{chieff}).

\begin{figure}[h!]
	\centering
	\includegraphics[width=0.6\linewidth]{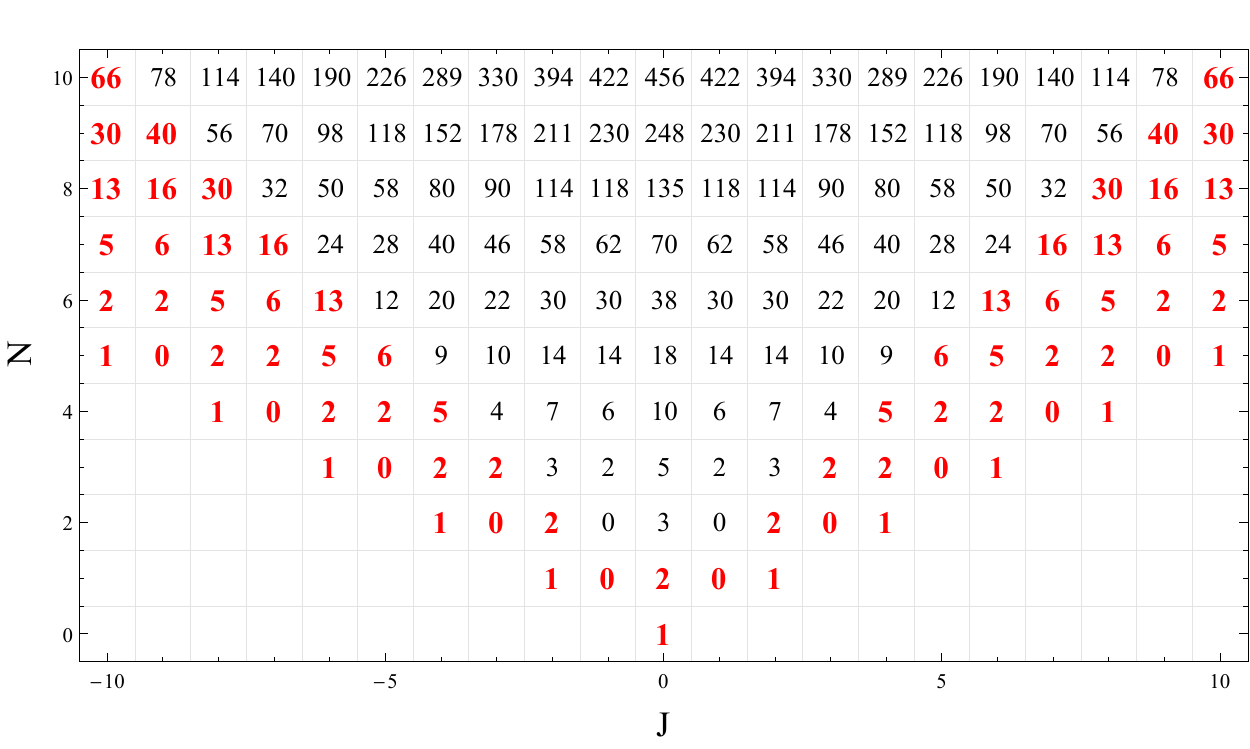}
	\includegraphics[width=0.6\linewidth]{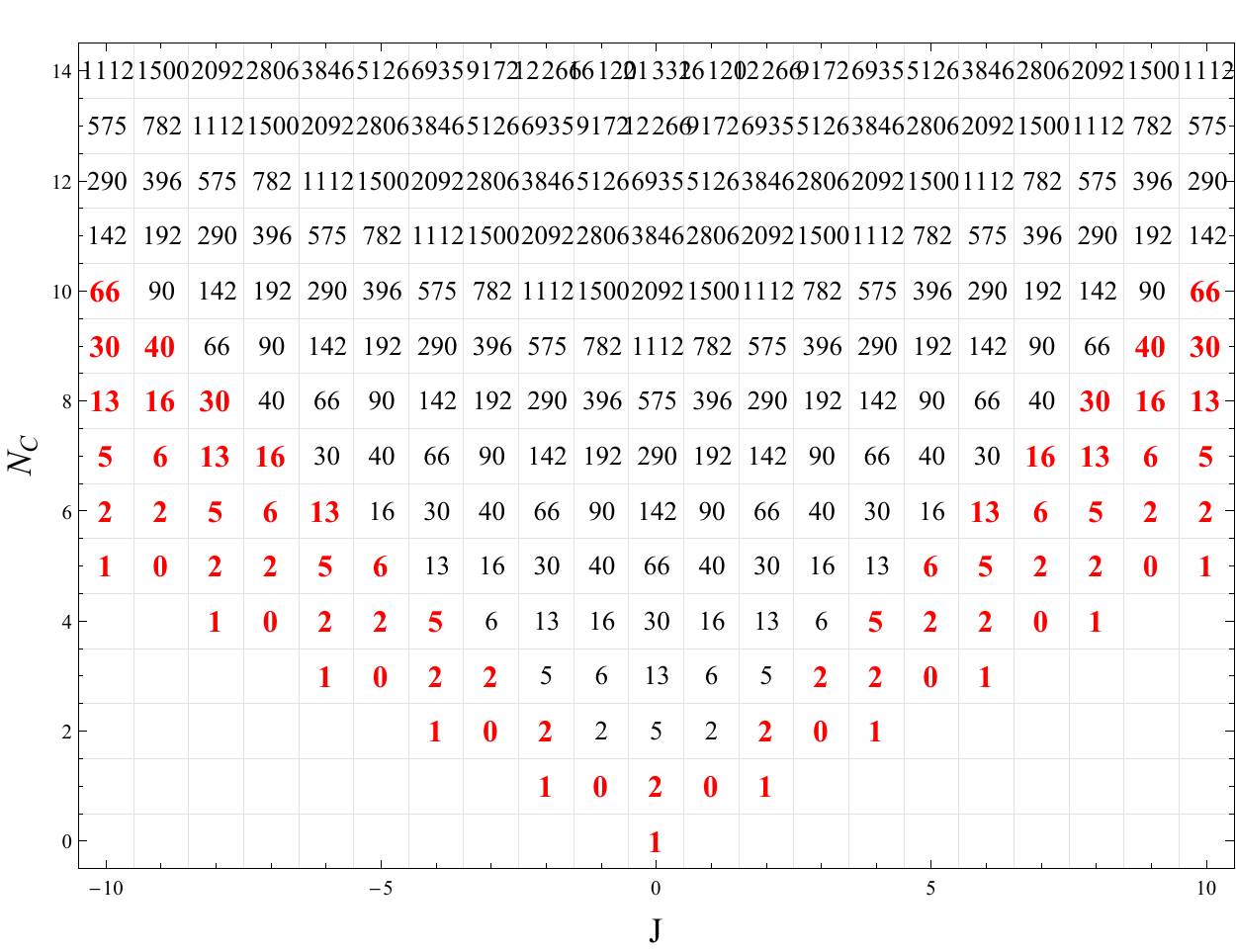}
	\caption{Multiplicities of states with angular momentum $J$ and level $N_c$ for the ASA spectrum (the top panel) and for effective closed strings (the bottom panel). Red entries are the same for both spectra.}
	\label{fig:effectrivestring}
\end{figure}

Clearly, we see that the ASA and effective string spectra are not the same for closed strings. We illustrate this in Fig.~\ref{fig:effectrivestring} where we presented the multiplicities of both 
spectra for a range of values of $J$ and $N$. Note that here we label closed strings levels by 
\[
N_c={N\over 2}\;
\]
rather than by $N$.

A couple of observations can be made by comparing the two spectra.
First, the ASA and effective strings agree in the vicinity of the leading Regge trajectory. This is similar to how the open string ASA spectrum (which, in our terminology, is the same as the effective open string spectrum) was found to agree with the light cone spectrum in  \cite{Dubovsky:2016cog}\footnote{Here, by light cone spectrum we mean the spectrum obtained using analytic continuation from $D>3$. Unlike the light cone spectrum of \cite{Mezincescu:2010yp}, this one is anyon free.}. On one side, this agreement had to be expected, because in the vicinity of the leading Regge tarjectory  effective string theory should be trustworthy. On the other hand it can be considered as an additional consistency check of the ASA, which relies on the ad hoc assumption (\ref{LRopen}). In Appendix D we prove that this agreement holds for all $N_c$ and $J$, satisfying $N_c\leq J\leq 2 N_c$. 

\begin{figure}
	\centering
	\includegraphics[width=0.7\linewidth]{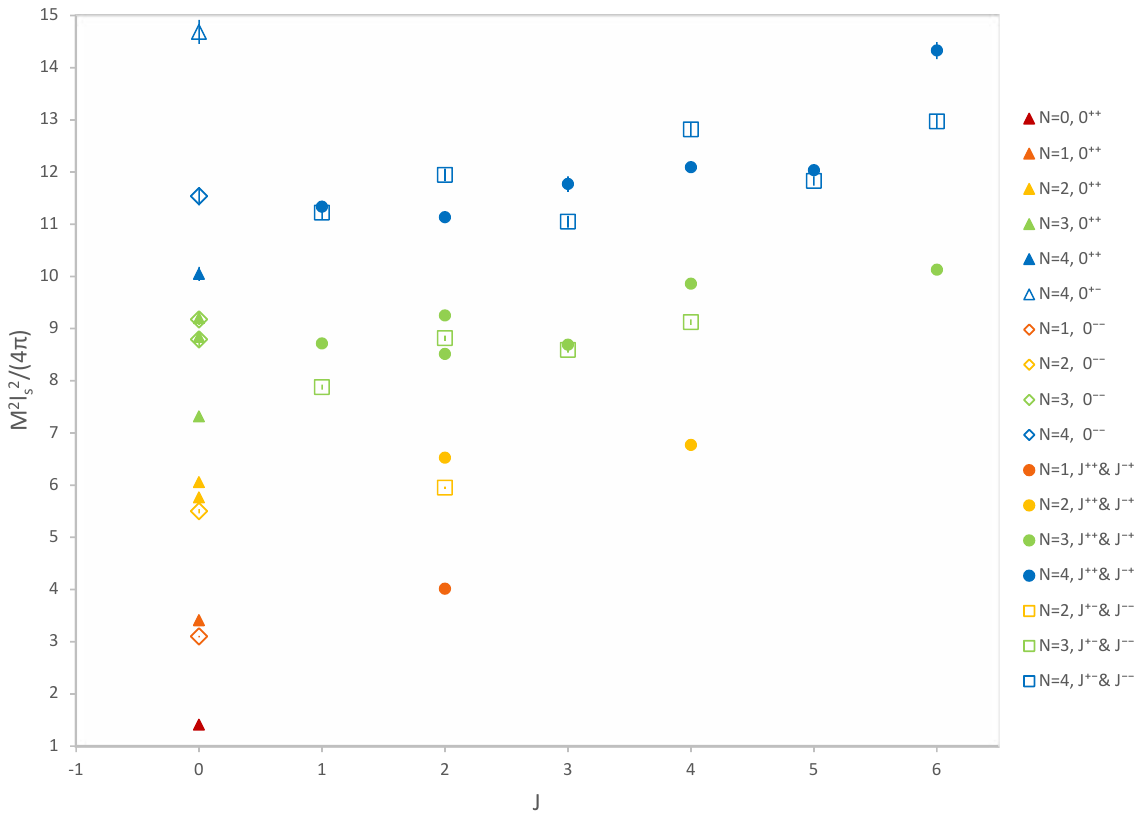}
	\caption{Lattice results for glueball masses and quantum numbers \cite{Conkey:2019blu}. For level $N=4$ we included one state only for each set of quantum numbers.}
	\label{fig:latticeglueballs}
\end{figure}

To illustrate which of the two spectra is in a better agreement with the actual glueball data, we present results of the glueball spin determinations from   \cite{Conkey:2019blu} in Figure~\ref{fig:latticeglueballs}. These lattice results are in excellent agreement for the states corresponding to $N_c=0,1,2,3$ ASA levels (39 states in total). As discussed in more details in  \cite{Conkey:2019blu}, lattice results broadly agree with the ASA spectrum also for 64 states corresponding to $N=4$ level, although more accurate simulations are needed to reach a definitive conclusion for these state. 

On the other hand, existing results demonstrate clear contradictions between effective closed strings and actual glueball spectra starting already from the $N=2$ level. Interestingly, when the two spectra disagree, effective strings always predict extra states compared to the ASA. This is somewhat surprising---naively one might expect that effective  theory would miss some states, rather than overpredict them. Perhaps, this is related to the constraints (\ref{poscond1}), which are ignored when extrapolating the effective theory results to arbitrary $J$ and $N$. However, for some reason, neglecting an analogous constraint (\ref{poscond}) for open strings in the ASA derivation does not lead to any contradictions (at least, as far as currently available lattice data is concerned).

Note also that, independently of lattice data, the effective theory spectrum extrapolated to all $J$ and $N$ is incompatible with the tensor square structure. Indeed,
the spectrum constructed in section \ref{sec:effective} does not pass even the very basic test---total multiplicities at fixed $N$ are
not always given by integer squared.   

Finally, in spite of the difference between the two spectra, they show the same exponential growth of the total number of states at large masses, which is determined by $P(x)^2$ factor. Namely, using the Hardy--Ramanujan asymptotics for the number partitions 
 \[
 P_n\sim e^{\pi\sqrt{2n\over 3}},
 \]
 we find that both spectra exhibit Hagedorn behavior with the Hagedorn temperature equal to
 \[
 T_H^{-1}=\ell_s\sqrt{\pi\over 3}\;.
 \]
 
\section{Conclusions and future directions}
\label{sec:concl}
To summarize, in this work we constructed a perturbative spectrum of closed effective strings in $D=3$, by expanding around a classical folded rotating rod solution with large spin $J_r$.
This spectrum is expected to be accurate in the vicinity of the leading Regge trajectory. Indeed, we found that it agrees there with the ASA spectrum and with the lattice Yang-Mills data. 
In particular, it correctly reproduces the angular momentum quantization of the glueball states on the leading Regge trajectory.

On one side, this agreement provides a consistency check for the ASA. On the other hand, an observation that the two spectra disagree for highly excited states 
makes it even more surprising that the ASA spectrum agrees so well with the lattice data. Indeed, the ASA spectrum is also based on the extrapolation of the high $J$ results to all values of $J$. As we see, for some reason though this extrapolation works better at the level of open strings.

Throughout the paper, we focused on counting the state multiplicities for different values of $N$ and $J$. In addition, the ASA spectrum predicts also $C$ parity assignments for string states, which also agree with the lattice data. It is straightforward to obtain the analogous prediction for the  closed effective strings. Namely, $C$ conjugation acts by changing an orientation of a closed string, {\it i.e.} with the closed effective string theory it corresponds to the worldsheet parity $\sigma\to-\sigma$. This transformation leaves the folded rotating rod solution (\ref{XClosed}) invariant and acts by exchanging the two towers of oscillators in (\ref{Nab}),
\[
C(a_n)=b_n\;.
\]
We checked that parity assignments agree between the ASA and closed effective strings when the two spectra predict the same multiplicities.

Assigning the $P$ parity is harder. Indeed, $P$ parity flips a sign of the angular momentum, so the classical background is not invariant under this transformation. A perturbative quantization is only consistent for excited states whose angular momentum is close to angular momentum of the classical background. As a result, the $P$ parity transformation necessarily maps one ``patch" covered by effective theory into another. For instance, in order to construct states with definite parity on the leading Regge trajectory one needs to take linear superpositions of rotating rods of opposite momenta. This makes it problematic to come up with $P$ parity assignment at the level of effective strings, either closed or open. 

Nevertheless, as explained in \cite{Dubovsky:2016cog}, the ASA allows one to obtain $P$ parity assignments for the majority of states. First, only spin 0 states need to be considered---all states with $J\neq 0$ automatically come in pairs of opposite parity. Furthermore, for the same reason, the only closed string states whose $P$ parity is left undetermined in the ASA are the $J=0$  states obtained as a tensor product of two open string $J=0$ states. The first such state appears at the $N_c=4$ level. Furthermore, this necessarily leads to the appearance of exotic $CP=-1$ states---and this is the only way such states can (and must) appear. Interestingly, a state with $CP=-1$ and $P=1$ has indeed been found in lattice simulations \cite{Conkey:2019blu}, with the mass in the correct ballpark (although, the mass determination for this highly excited state is quite imprecise). It will be interesting to see whether the ASA can be extended to predict the $P$ parity of these exotic  states.

Our main motivation for performing the semiclassical analysis directly in the closed string level is to use this as a starting point to calculate mass splittings, coming from either higher orders in $1/J$ expansion or from higher dimensional operators. This calculation is hard to implement within the ASA framework because it is unclear how to determine interactions between the two components of the tensor square. This problem does not arise for effective closed string calculations. However, as our results show, this framework allows to calculate masses only for a limited set of states. Still, it will be interesting to perform this calculation and to compare the results with the lattice spectrum. To implement this one needs to overcome 
the challenge of how to treat the fold singularities of the rotating rod solution.

Finally,  another  avenue for future work is to apply these ideas to $D=4$. This is particularly interesting given that the recent lattice simulations \cite{athenodorou2020glueball} provide a new wealth of high quality data to test against stringy predictions. The major novelty when moving to the $D=4$ case is the presence of a massive
excitation on the worldsheet---the worldsheet axion \cite{Dubovsky:2014fma,Dubovsky:2015zey}. 

{\it Acknowledgements.} This work is supported in part by the NSF award PHY-1915219, by the BSF grant 2018068 and by the CSIC I+D grant number 583.

\section*{Appendix A: Details of the gauge fixing for open strings}
The action (\ref{fo}) includes two first class Virasoro constraints
\begin{gather}
V_\tau= \ell_s^2\Pi^\mu \Pi_\mu+\ell_s^{-2}\d_\sigma X^\mu
\d_\sigma X_\mu\;,\\
V_\sigma=\Pi_\mu\d_\sigma X^\mu\;.
\end{gather}
As usual these generate gauge transformations of any physical observable ${\cal O}$ via Poisson brackets,
\be
\delta{\cal O}=\left\{\int d\sigma \l \xi^\tau(\tau,\sigma)V_\tau+\xi^\sigma(\tau,\sigma)V_\sigma\r ,{\cal O}\right\}\;.
\ee
As a result, embedding coordinates and momenta of the string transform as 
\begin{gather}
\label{deltaX}
\delta X^\mu=\ell_s^2\xi^\tau\Pi^\mu+\xi^\sigma\d_\sigma X^\mu\\
\label{deltaP}
\delta\Pi^\mu=\ell_s^{-2}\d_\sigma\l\xi^\tau\d_\sigma X^\mu\r+\d_\sigma(\Pi^\mu\xi^\sigma)\;.
\end{gather}
As explained in \cite{Brink:1988nh}, these transformations reduce to the conventional world-sheet reparametrizations on-shell.

To impose the static gauge conditions we first make use of the $V_\tau$ generator to fix $X^0$ to be of the form (\ref{X0}). The resulting $X^0$ is invariant under the $V_\sigma$ generator, and the latter can be used to  fix 
$\Pi^0$ in the form (\ref{P0}). Note that, as follows from (\ref{deltaP}),  the total momenta $\int d\sigma \Pi^\mu$ are invariant under gauge transformations (as expected), so we cannot fix $\Pi^0$ to be an arbitrary constant, and rather need to keep it as a general function of $\tau$ as we did in (\ref{P0}).
 

\section*{Appendix B: Solving the linearized Virasoro constraints for open strings}
By expanding the Virasoro constraints to linear order in perturbations for open strings and performing the Fourier decomposition, one arrives at the following set of equations
\begin{gather}
A_n=C_{n+2}\\
B_n=D_{n+2}
\end{gather}
where $n=0,1,\dots$, and 
\be
A_n=\left\{
\begin{array}{c}
\ell_s^2\mbox{\rm Re}\,\pi_n+n\,\mbox{\rm Im}\,\chi_n\,,\; n>0\\
\ell_s^2\sqrt{2\over \pi}\,\mbox{\rm Re}\,p\,,\; n=0
\end{array}
\right.
\ee   
\be
B_n=\left\{
\begin{array}{c}
n\,\mbox{\rm Re}\,\chi_n-\ell_s^2\,\mbox{\rm Im}\,\pi_n\,,\; n>0\\
-\ell_s^2\sqrt{2\over \pi}\,\mbox{\rm Im}\,p\,,\; n=0
\end{array}
\right.
\ee
\begin{gather}
C_n=\ell_s^2\,\mbox{\rm Re}\,\pi_n-n\,\mbox{\rm Im}\,\chi_n\,,\; n\geq 2\\
D_n=n\,\mbox{\rm Re}\,\chi_n+\ell_s^2\,\mbox{\rm Im}\,\pi_n\,,\; n\geq 2\;.
\end{gather}
One more linearized constraint reads
\be
\label{VirE}
E-E_r=\sqrt{\pi\over 2}\l \mbox{\rm Im}\,\pi_1+\ell_s^{-2}\chi_1\r\;.
\ee
Taking into account the expression (\ref{J}) for the total angular momentum, this last constraint (\ref{VirE}) can be rewritten as
\[
E-E_r=\ell_s^{-1}\sqrt{\pi\over 2J_r}\l J-J_r\r\;,
\]
which is simply a local linear approximation to the shape of the leading Regge trajectory (\ref{Regge}).

We make use of these constraints to express the energy $E$ and all $C_n$, $D_n$ variables in terms of the remaining physical phase space variables ($x$, $p$, $\delta$, $J$, $A_n$, $B_n$) with $n\geq 1$. By plugging the result into the canonical one-form (\ref{canonical}) we arrive at the following expression for the reduced canonical structure (up to a total derivative)
\be
\label{redO}
\Omega_{red}= p_i\d_\tau \tilde{x}^i
+ J\d_\tau \delta_c+\sum_{n=1}^\infty P_n\d_\tau Q_n\;,
\ee
where we defined the following canonical variables on the reduced phase space,
\begin{gather}
\label{shift}
 \tilde{x} =  x -{i\ell_s^2\over 4\pi}p\\
\delta_c=\delta-{A_1\over 2\ell_s\sqrt{J_r}}\\
Q_n={1\over (n(n+2))^{1/2}}B_n\\
P_n={n+1\over\ell_s^2 (n(n+2))^{1/2}}A_n
\end{gather}
To calculate the full quadratic action and the mass spectrum of open strings we need to evaluate the constraint (\ref{VirE}) to the second order in perturbations (because the energy $E$ enters linearly in the action).
In fact, the resulting mass shell condition looks the simplest without performing any expansion,
\be
\label{masscond}
E^2=|p|^2+{2\pi\over \ell_s^2}\l 
J+\mbox{\rm Im }\l p^*x + \sum_{n=1}^\infty\pi_n^*\chi_n\r+{1\over 2}\sum_{n=1}^\infty\l\ell_s^2|\pi_n|^2+\ell_s^{-2}n^2|\chi_n|^2 \r\r\;.
\ee
To deduce the leading order physical spectrum we need to rewrite (\ref{masscond}) in terms of the canonical variables on the reduced phase
space. This results in the following expression for the energy,
\be
E^2={|p|^2}+{2\pi\over\ell_s^2}\l J + \mbox{\rm Im }\l p^*\tilde{x} \r + \,{\ell_s^2\over 2}\sum_{n=1}^\infty \l P_n^2+{(n+1)^2\over\ell_s^4}Q_n^2\r\r.
\ee

\noindent The quadratic Hamiltonian for open strings in the reduced phase space reads

\begin{equation}
H =  \ell_s\sqrt{\frac{2 J_r}{\pi}}E - J.
\end{equation} 

\noindent From these expressions it is clear that $J$ is conserved, given that its conjugate variable $\delta_c$ does not appear in the Hamiltonian. Furthermore, it is clear that $J$ is quantized in integer units since $\delta_c$ is  $2\pi$ periodic. 

 However, the conservation of neither $|p|^2$ nor of the center of mass angular momentum is manifest yet.
These conservations can be shown by performing the change of variables
\begin{align}
x_c = e^{i(\tau+\delta_c)}\tilde{x}\label{transfphysx}\\
p_c = e^{i(\tau+\delta_c)}p,\label{transfphysp}
\end{align} 
for which clearly $|p|^2 = |p_c|^2$ and $\mbox{\rm Im}\left(\tilde{x}^*p\right)=\mbox{\rm Im}\left(x_c^*p_c\right)$. After this change of variables, the canonically conjugate variable to $\delta_c$ is the internal angular momentum
\begin{equation}
\label{internalI}
I = J - \mbox{\rm Im}\left(x_c^*p_c\right).
\end{equation}
 Notice that under this time-dependent transformation the Hamiltonian also changes and becomes
\begin{equation}
H =  \ell_s\sqrt{\frac{2 J_r}{\pi}}E - I.
\end{equation} 
In these variables it is immediate that $I$ is quantized in integer units and conserved by the same argument as before for $J$. This then implies that the orbital angular momentum $\mbox{\rm Im}\left(x_c^*p_c\right)$ is also conserved and quantized. Furthermore, using $I$ as a canonical variable we see that $x_c$ no longer appears explicitly in the Hamiltonian, and thus $p_c$ is conserved.

 To summarize, the reduced phase space may be parametrized by the canonical pairs $\left\{x_c,p_c\right\}$, $\left\{\delta_c,I\right\}$ and $\left\{Q_n,P_n\right\}_{n>0}$. Up to quadratic order in the perturbations the energy is given by  

\be
E^2={|p|^2}+{2\pi\over\ell_s^2}\l I  + \,{\ell_s^2\over 2}\sum_{n=1}^\infty \l P_n^2+{(n+1)^2\over\ell_s^4}Q_n^2\r\r.
\ee
 which reduces to  \eqref{energy_vs_N_open} by switching to the creation/annihilation variables.

\section*{Appendix C: Solving the linearized Virasoro constraints for closed strings}
Fourier decomposition  of the linearized Virasoro constraints for closed strings results in the following set of equations
\begin{gather}
A_n=C'_{n+2} \\
B_n=D'_{n+2} \\
A'_n=C_{n+2} \\
B'_n=D_{n+2}\\
\label{VirEClosed0}
\,\mbox{\rm }\,\chi_1+\ell_s^2\,\mbox{\rm Im}\,(\pi_1)=\chi_{-1}+\ell_s^2\,\mbox{\rm Im}\,(\pi_{-1})\;,
\end{gather}
where $n=0,1,\dots$, and 
\be
A_n=\left\{
\begin{array}{c}
\ell_s^2\mbox{\rm Re}\,\pi_n+n\,\mbox{\rm Im}\,\chi_n\,,\; n>0\\
{\ell_s^2\over 2}\sqrt{2\over \pi}\,\mbox{\rm Re}\, p\,,\; n=0
\end{array}
\right.
\ee
\be
B_n=\left\{
\begin{array}{c}
n\,\mbox{\rm Re}\,\chi_n-\ell_s^2\,\mbox{\rm Im}\,\pi_n\,,\; n>0\\
-{\ell_s^2\over 2}\sqrt{2\over \pi}\,\mbox{\rm Im}\, p\,,\; n=0
\end{array}
\right.
\ee
\begin{gather}
C_n=\ell_s^2\,\mbox{\rm Re}\,\pi_n-n\,\mbox{\rm Im}\,\chi_n\,,\; n\geq 2\\
D_n=n\,\mbox{\rm Re}\,\chi_n+\ell_s^2\,\mbox{\rm Im}\,\pi_n\,,\; n\geq 2\;
\end{gather}
\be
A'_n=\left\{
\begin{array}{c}
\ell_s^2\mbox{\rm Re}\,\pi_{-n}+n\,\mbox{\rm Im}\,\chi_{-n}\,,\; n>0\\
{\ell_s^2\over 2} \sqrt{2\over \pi}\,\mbox{\rm Re}\, p\,,\; n=0
\end{array}
\right.
\ee
\be
B'_n=\left\{
\begin{array}{c}
n\,\mbox{\rm Re}\,\chi_{-n}-\ell_s^2\,\mbox{\rm Im}\,\pi_{-n}\,,\; n>0\\
-{\ell_s^2\over 2} \sqrt{2\over \pi}\,\mbox{\rm Im}\, p\,,\; n=0
\end{array}
\right.
\ee
\begin{gather}
C'_n=
\ell_s^2\mbox{\rm Re}\,\pi_{-n}-n\,\mbox{\rm Im}\,\chi_{-n}\,,\; n\geq 2\\
D'_n=
n\,\mbox{\rm Re}\,\chi_{-n}+\ell_s^2\,\mbox{\rm Im}\,\pi_{-n}\,,\; n\geq 2\;.
\end{gather}
Yet another constraint reads
\[
E-E_r=\sqrt{\pi\over 2}\l \ell_s^{-2}(\chi_1+\chi_{-1})+\mbox{\rm Im}\,(\pi_1+\pi_{-1})\r\;.
\]
 As a consequence of (\ref{VirEClosed0}), this reduces to
\be
\label{VirEClosed}
E-E_r=\sqrt{2\pi}\l \ell_s^{-2}\chi_1+\mbox{\rm Im}\,\pi_1\r\;.
\ee
Given the expression (\ref{JClosed}) for the total angular momentum, (\ref{VirEClosed}) can be rewritten as 
\[
E-E_r=\ell_s^{-1}\sqrt{\pi\over J_r}\l J-J_r\r\;,
\]
which is simply a local linear approximation to the shape of the leading Regge trajectory (\ref{ReggeClosed}).

We follow the same approach as for open strings to express the energy $E$ and all $C_n$, $D_n$, $C'_n$, $D'_n$ variables in terms of the remaining physical phase space variables ($x$, $p$, $\delta$, $J$, $A_n$, $B_n$, $A'_n$, $B'_n$) with $n\geq 1$. By plugging the result into the canonical one-form (\ref{canonicalClosed}), one arrives at the following  reduced canonical structure (up to a total derivative)
\be
\label{redOClosed}
\Omega_{red}= p_i\d_\tau \tilde{x}^i
+ J\d_\tau \delta_c+\sum_{n=1}^\infty (P_n\d_\tau Q_n + P'_n\d_\tau Q'_n)\;,
\ee
where we defined the following canonical variables on the reduced phase space,
\begin{gather}
\label{shiftClosed}
\tilde x= x -\frac{i\ell_s^2}{8\pi}p\\
\delta_c=\delta-{(A_1+A'_1)\over 2\ell_s\sqrt{2J_r}}\\
Q_n={1\over (n(n+2))^{1/2}}B_n\\
P_n={n+1\over\ell_s^2 (n(n+2))^{1/2}}A_n\\
Q'_n={1\over (n(n+2))^{1/2}}B'_n\\
P'_n={n+1\over\ell_s^2 (n(n+2))^{1/2}}A'_n
\end{gather}

The full quadratic action and the mass spectrum of closed strings can be evaluated by the constraint (\ref{VirEClosed}) to the second order in perturbations (because the energy $E$ enters linearly in the action).
Without performing any expansion, the mass shell condition will be 
\be
\label{masscondClosed}
E^2=|p|^2+{4\pi\over \ell_s^2}\l 
J +\mbox{\rm Im }\l p^*x + \sum_{n\neq0}\pi_n^*\chi_n\r+{1\over 2}\sum_{n\neq0}\l\ell_s^2|\pi_n|^2+\ell_s^{-2}n^2|\chi_n|^2 \r\r\;.
\ee
Rewriting (\ref{masscondClosed}) in terms of the canonical variables on the reduced phase space, one arrives at the following leading order physical spectrum 
%
%
\be
E^2={|p|^2}+{4\pi\over\ell_s^2}\l J+ \mbox{\rm Im }\l p^*\tilde{x} \r+\,{\ell_s^2\over 2}\sum_{n=1}^\infty \l P_n^2+{(n+1)^2\over\ell_s^4}Q_n^2{+P'}_n^2+{(n+1)^2\over\ell_s^4}{Q'}_n^{2}\r\r.
\ee
 The quadratic Hamiltonian for closed strings in the reduced phase space reads

\begin{equation}
H =  \ell_s\sqrt{\frac{J_r}{\pi}}E - J.
\end{equation} 
 From here an argument identical to that performed for open strings at the end of Appendix B yields formula \eqref{energy_vs_N_closed} for the ``almost physical'' spectrum of closed strings at the quadratic order in perturbations.

\section*{Appendix D: Some algebra with generating functions}

We start from the ASA prescription for open string states, whose degeneracies are encoded in the generating function

\begin{equation}
\chi_{open} = \sum_{J\in \mathbb{Z}}x^{|J|}(1-x)P(x)e^{iJ\theta},
\end{equation}
where $P(x)$ denotes the Euler generating function
\begin{equation}
P(x) = \prod_{m=1}^{\infty}(1-x^m)^{-1} = \sum_{m=0}^{\infty}x^mP_m.
\end{equation}
In order to construct the closed string ASA generating function, we need to extract the spin content from $\chi_{open}$ at each level. Simple manipulations yield

\begin{equation}
\chi_{open} = 1 + \sum_{N=1}^{\infty}x^N\left(\sum_{J=-N}^{N}P_{N-|J|}e^{iJ\theta} - \sum_{J=-(N-1)}^{N-1}P_{N-|J|-1}e^{iJ\theta}\right).
\end{equation}
From here, the ASA level matching prescription implies that the ASA closed string generating function is then
\begin{equation}
\label{chiclosed}
\chi_{cl} = 1 + \sum_{N=1}^{\infty}x^N\left(\sum_{J=-N}^{N}P_{N-|J|}e^{iJ\theta} - \sum_{J=-(N-1)}^{N-1}P_{N-|J|-1}e^{iJ\theta}\right)^2.
\end{equation}
Our objective is to extract the coefficients $\chi_{cl}(N,L)$ from here. These are defined as
\begin{equation}
\chi_{cl} = 1 + \sum_{N=1}^{\infty}\sum_{N\in \mathbb{Z}}x^Ne^{iL\theta}\chi_{cl}(N,L).
\end{equation}
By taking the square in (\ref{chiclosed})  it can be shown that 
\begin{gather}
\chi_{cl}(N,2N) = P_0^2 =1\\
\chi_{cl}(N,2N-1) = 2P_1P_0-2P_0^2=0
\end{gather}
 and for $0<L\leq 2N-2$
\begin{align}
\chi_{cl}(N,L) &= \sum_{J=L-N}^{N}P_{N-|J|}P_{N-|L-J|} + \sum_{J=L-(N-1)}^{N-1}P_{N-1-|J|}P_{N-1-|L-J|}\nonumber\\ &- 2\sum_{J=L-(N-1)}^{N}P_{N-|J|}P_{N-1-|L-J|}.
\end{align}  
This last expression simplifies considerably close to the leading Regge trajectory. Namely, if we take $L = 2N-a$, then in the range $a\leq N$ we have
\begin{equation}
\chi_{cl}(N,2N-a) = \sum_{l=0}^{a}P_lP_{a-l} + \sum_{l=0}^{a-2}P_lP_{a-l-2} -2\sum_{l=0}^{a-1}P_lP_{a-l-1}. 
\end{equation}

On the other hand, before the $Z_2$ projection, the closed effective string generating function corresponds to the one generated by the two towers of oscillators $a_n$ and $b_n$ described in equation \eqref{Nab}. Thus it is given by
\begin{equation}
\chi_{eff} = \sum_{L=-\infty}^{\infty}e^{iL\theta}x^{|L|}(1-x)^2P(x)^2,
\end{equation}
 which can be written as
\begin{align}
\chi_{eff} &= \sum_{N=0}^{\infty}\sum_{J=-N}^{N}\sum_{M=0}^{N-|J|}x^Ne^{iJ\theta}P_M P_{N-M-|J|}\nonumber\\
&+\sum_{N=2}^{\infty}\sum_{J=-(N-2)}^{N-2}\sum_{M=0}^{N-|J|-2}x^Ne^{iJ\theta}P_M P_{N-M-|J|-2}\nonumber\\
&-2\sum_{N=1}^{\infty}\sum_{J=-(N-1)}^{N-1}\sum_{M=0}^{N-|J|-1}x^Ne^{iJ\theta}P_M P_{N-M-|J|-1}.
\label{chieff2}
\end{align}
In order to implement the $Z_2$ projection and get the generating function corresponding to the physical Hilbert space, we need to keep only even powers of $N$ in expression \eqref{chieff2}. After some manipulations we obtain
\begin{align}
\chi^{Z2}_{eff} &= 1 + \sum_{N=1}^{\infty}x^Ne^{2Ni\theta}+\sum_{N=1}^{\infty}\sum_{J=-2(N-1)}^{2(N-1)}x^Ne^{iJ\theta}\left(\sum_{l=0}^{2N-|J|}P_lP_{2N-|J|-l}\right.\nonumber\\
&\left.+\sum_{l=0}^{2N-|J|-2}P_lP_{2N-|J|-l-2}-2\sum_{l=0}^{2N-|J|-1}P_lP_{2N-|J|-l-1}\right).
\end{align}
Defining $\chi^{Z2}_{eff}(N,L)$ analogously by
\begin{equation}
\chi^{Z2}_{eff} = \sum_{L=-\infty}^{\infty}\sum_{N=0}^{\infty}e^{iL\theta}x^N\chi^{Z2}_{eff}(N,L),
\end{equation}
it is straightforward to check that 
\begin{equation}
\chi_{cl}(N,2N-a) =\chi^{Z2}_{eff}(N,2N-a)
\end{equation}
in the range $0\leq a \leq N$.

\bibliographystyle{utphys}
\bibliography{dlrrefs}
\end{document}